# Mathematical Estimation of Logical Masking Capability of Majority/Minority Gates Used in Nanoelectronic Circuits

P. Balasubramanian
School of Electrical and Electronic Engineering
Nanyang Technological University
Singapore
e-mail: balasubramanian@ntu.edu.sg

R. T. Naayagi
School of Electrical and Electronic Engineering
Newcastle University International Singapore
Singapore
e-mail: naayagi.ramasamy@ncl.ac.uk

*Abstract*—In nanoelectronic circuit synthesis, the majority gate and the inverter form the basic combinational logic primitives. This paper deduces the mathematical formulae to estimate the logical masking capability of majority gates, which are used extensively in nanoelectronic digital circuit synthesis. The mathematical formulae derived to evaluate the logical masking capability of majority gates holds well for minority gates, and a comparison with the logical masking capability of conventional gates such as NOT, AND/NAND, OR/NOR, and XOR/XNOR is provided. It is inferred from this research work that the logical masking capability of majority/minority gates is similar to that of XOR/XNOR gates, and with an increase of fan-in the logical masking capability of majority/minority gates also increases.

*Keywords-logic gates; majority gate; minority gate; logical masking; fault tolerance; nanoelectronics*

## I. Introduction

Emerging nanotechnologies for beyond-CMOS futuristic electronic designs include quantum cellular automata [1]–[5], spin-based devices [6]–[11], resistive random access memories [12]–[16], biological circuits [17], [18] etc. These technologies share a common ground in that they widely use the majority gate as the basic building block. Although the logic realization of combinational circuits using the 3-input majority gate was proposed as early as the 1960's [19], [20], the latest nanometric technologies [21] have given rise to a renewed interest in synthesizing digital circuits using the majority gate and the inverter. Several digital electronic circuit synthesis and optimization techniques [22]–[27] have been proposed of late which utilize only the majority gate and the inverter as the basic building blocks. Also, logic synthesis schemes utilizing the minority gate and the inverter were proposed [23], [28], with the minority logic function being the complement of the majority logic function.

References [29]–[32] analyzed the impact of probabilistic input errors on the outputs of some logic gates and some logic circuits. Reference [33] analyzed the logical masking capability of conventional gates such as NOT (i.e. inverter), AND/NAND, OR/NOR, and XOR/XNOR, where logical masking refers to the inherently ability of a gate to tolerate fault(s) on its input(s) and still produce the correct output. Note that not all the fault(s) which may occur on the input(s) of a gate, which may be temporary or permanent in nature, would result in error(s) on its output. For example, in a 2-input AND gate if its inputs are 0 and 1, then its output is 0. Now supposing that its input which is 1 has become 0 due to a fault, the output of the AND gate would still retain the correct value of 0.

In this paper, we specifically analyze the logical masking capability of majority/minority gates (or functions) which are extensively used in nanoelectronic designs [21]. Moreover, the majority logic function is widely employed [34]–[37] in redundancy architectures which are indeed commonplace in many mission-critical and safety-critical electronic circuit and system designs [38]–[41]. Further, the information about the logical masking capability of a gate may be useful to design digital circuits with improved intrinsic fault tolerance [42]. This tends to assume significance in the context of emerging nanoelectronic designs which are more likely to experience temporary or permanent faults or failures due to radiation and other phenomena [43]–[45].

The rest of this paper is organized as follows. Section 2 describes the generalized error metrics which are used to estimate the logical masking capabilities of various gates including the majority and minority gates. The transistor level circuits of the 3-input majority and minority gates are also presented. Section 3 presents a comparison of the logical masking capabilities of commonly used gates viz. NOT, AND/NAND, OR/NOR, XOR/XNOR, and the majority and minority gates based on the general error metrics. Finally, the conclusions are stated in Section 4.

## II. Majority/Minority Gates and Their Generalized Error Metrics

Let us consider for an example illustration the truth table of the 3-input majority and minority functions given below. The inputs and outputs feature binary data (i.e. 0 and 1).

TABLE I. Truth Table of 3-Input Majority and Minority Gates (Logic Functions)

| Primary Inputs | | | Majority Output | Minority Output |
|---|---|---|---|---|
| A | B | C | Z | Y |
| 0 | 0 | 0 | 0 | 1 |
| 0 | 0 | 1 | 0 | 1 |
| 0 | 1 | 0 | 0 | 1 |
| 0 | 1 | 1 | 1 | 0 |
| 1 | 0 | 0 | 0 | 1 |
| 1 | 0 | 1 | 1 | 0 |
| 1 | 1 | 0 | 1 | 0 |
| 1 | 1 | 1 | 1 | 0 |

In Table I, A, B and C are the primary inputs, Z is the majority function output, and Y is the minority function output. Whenever a majority of the inputs i.e. at least 2 out of 3 inputs in Table I is 1, the majority function output Z will be 1, otherwise it will be 0. On the contrary, whenever a minority of the inputs i.e. if at least 2 out of 3 inputs in Table I is 0, the minority function output Y will be 1, otherwise it will be 0. Thus, the majority and the minority function outputs are complementary to each other.

In the 3-input majority logic function, at least 2 out of the 3 inputs should be 1. Likewise, in the 5-input majority logic function, at least 3 out of the 5 inputs should be 1, and so on. Generalizing this, we note that in an $n$-input majority logic function, at least $(n + 1)/2$ out of the $n$ inputs should be 1, and in an $n$-input minority logic function, at least $(n + 1)/2$ out of the $n$ inputs should be 0 for the corresponding function outputs to be 1.

The sum-of-products and product-of-sums expressions for the 3-input majority function are given by (1) and (2). The complex gates AO222 and OA222 of a standard digital cell library [46], which directly synthesize (1) and (2) are shown in Figure 1. The respective transistor level circuits corresponding to the static CMOS style logic implementation are also shown.

$$Z_{SOP} = AB + BC + AC = AB + C(A + B) \quad (1)$$

$$Z_{POS} = (A + B)(B + C)(A + C) = (A + B)(AB + C) \quad (2)$$

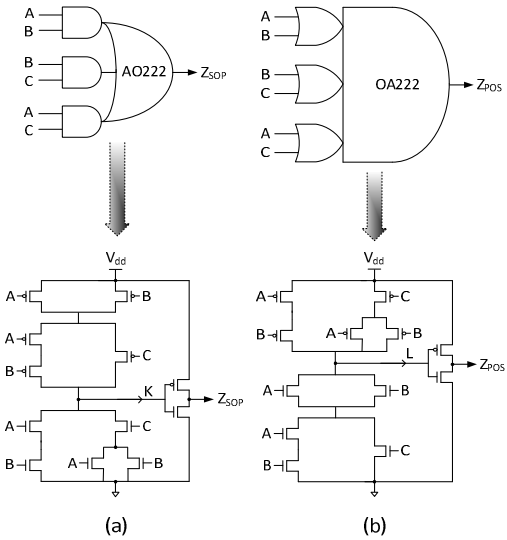

Notes:
(i) In (a), intermediate output K and primary output Z are complementary to each other. Likewise in (b), L and Z are complementary to each other
(ii) K is the inverted output of AO222 gate (i.e. AOI222 gate output)
(iii) L is the inverted output of OA222 gate (i.e. OAI222 gate output)
(iv) AO222 and OA222 are 'majority gates'; AOI222 and OAI222 are 'minority gates'

Figure 1. Two complex gate realizations of the 3-input majority function Z shown in Table I. The static CMOS style implementations are also shown. The intermediate outputs K and L in (a) and (b) represent the 3-input minority function outputs.

We define two generalized error metrics (GEMs) to estimate the logical masking capability of logic gates. GEMNIF, given below, was referred to as GEM in [33].

- GEMNIF is specified as the ratio of the total number of gate output errors to the total number of individual input faults while considering the application of all possible unique input patterns and their associated distinct faulty input patterns
- GEMFIC is given by the ratio of total number of gate output errors to the total number of faulty inputs while considering the application of all possible unique input patterns and their associated distinct faulty input patterns

Although GEMNIF and GEMFIC assume a uniform inputs distribution, as per the definitions, nevertheless the definitions may be modified to suit any practical scenario once the complete information about the input patterns applied and the subsequent input faults which have occurred are ascertained through an automatic test pattern generator or an automatic test equipment. Hence we leave it to the choice of a circuit designer to choose either or both metrics viz. GEMNIF and/or GEMFIC to quantitatively estimate the logical masking capabilities of logic gates as desired. Among these, possibly GEMNIF is computationally more expensive to determine but more accurate than GEMFIC.

*A. Derivation of GEMNIF Equation for Majority/Minority Gates*

The ON-set of a Boolean function comprises those input patterns which result in the function outputting 1, and the OFF-set of a Boolean function comprises those input patterns which result in the function outputting 0. Referring to Table I we find that the ON-set of Z comprises four elements viz. (011, 101, 110, 111), and the OFF-set of Z also comprises four elements viz. (000, 001, 010, 100). In Table I, the cardinalities of the ON-set and the OFF-set of Z are the same, which equals 4. In general, any $n$-input majority logic function will have identical ON-set and OFF-set cardinalities of magnitude $2^{n-1}$.

Whenever the inputs of a logic gate or a logic function migrate from the ON-set to OFF-set or vice-versa due to any fault occurrences, the corresponding gate (or function) output would be corrupted. For example, consider the binary input pattern 011 given in Table I, which reflects the values of inputs A = 0 and B = C = 1 for which Z = 1. Now supposing due to any fault(s) on these inputs, the assumed input pattern 011 changes to 000, 001, 010, or 100, the output Z would become 0, which is an error. This is because the faulty input patterns 000, 001, 010, or 100 correspond to the OFF-set of Z. On the other hand, even if due to any potential fault occurrence(s) on the inputs, supposing the assumed input pattern 011 changes to 101, 110, or 111, Z would still maintain the correct value of 1 since these faulty input patterns correspond to the ON-set of Z.

Let us once again consider the presumed input pattern of 011 whose faulty counterparts are 000, 001, 010 and 100. Between 011 and 000, there is a Hamming distance of 2, which implies the occurrence of two input faults. Likewise, between 011 and 001, 010, 100, the respective Hamming distances are 1, 1, and 3, which imply the occurrence of 1 fault, 1 fault, and 3 faults respectively on the inputs. The maximum Hamming distance is 3, and the minimum Hamming distance is 1. In general, for an $n$-input majority logic function, the maximum Hamming distance between an ON-set element and an OFF-set element would be $n$, and the minimum Hamming distance would be 1.

The total number of input faults which may occur on an ON-set element, at say different time instances, which may cause a migration of the input pattern to the corresponding OFF-set of an $n$-input majority logic function (or even any Boolean function) would be numerically expressed by $\sum_{k=1}^{n}\binom{n}{k}\times k$ in general, based on the principle of mathematical induction [33], where $\binom{n}{k} = nC_k$ and $k$ denotes the number of input faults which may occur ranging from 1 (i.e. the minimum Hamming distance) to $n$ (i.e. the maximum Hamming distance). Since there is a total of $2^n$ input patterns corresponding to an $n$-input logic gate or function, the total number of potential input fault(s) which may occur corresponding to the $n$-inputs would be given by $2^n\left\{\sum_{k=1}^{n}\binom{n}{k}\times k\right\}$ [33], which represents the denominator of GEMNIF corresponding to an $n$-input majority gate/function.

Since the ON-set and OFF-set cardinalities of an $n$-input majority gate or function are identical, for each input pattern in the ON-set there exists $2^{n-1}$ OFF-set elements which would cause an output error. Similarly, considering the migration of input patterns from the OFF-set to ON-set, there exists an equal number of ON-set elements. Hence, the total number of potential output errors which may be likely would be given by $2^{2n-1}$, which represents the numerator of GEMNIF for an $n$-input majority logic gate or function. Hence, GEMNIF for an $n$-input majority (or minority) logic gate or function is expressed by (3). The GEMNIF equations corresponding to NOT, AND, NAND, OR, NOR, XOR and XNOR are given in [33].

$$\text{GEMNIF}_{\text{Majority/Minority}} = \frac{2^{2n-1}}{2^n\left\{\sum_{k=1}^{n}\binom{n}{k}\times k\right\}} \quad (3)$$

Equation (3) holds well for any majority or minority logic gate or function. This is because the ON-set and OFF-set cardinalities of the majority or minority function outputs are the same, and just their outputs are reversed since the majority and minority functions are complementary to each other, as seen in Table I. In fact, (3) is similar to the GEMNIF equation derived for the XOR and XNOR gates in [33]. This is due to the reason that the XOR and XNOR gates are also equivalent gates [47], [48] as they also feature equal ON-set and OFF-set cardinalities like the majority and minority functions discussed here.

B. *Derivation of GEMFIC Equations for Different Gate Types*

As per the definitions stated earlier for GEMNIF and GEMFIC, it may be noted that their numerator components are the same but their denominator components are different. Given that any $n$-input Boolean (logic) function would have $2^n$ unique input patterns, hence for any unique input pattern there would be a total of $(2^n - 1)$ associated distinct faulty input patterns. Hence, the total number of distinct faulty input patterns with respect to all the unique input patterns of an $n$-input logic function would be specified by $2^n(2^n - 1)$. After substitution of the numerator components for GEMFIC from [33] for the different gate types, which are the same as those of GEMNIF, we obtain the GEMFIC expressions given below corresponding to NOT, AND/NAND/OR/NOR, and XOR/XNOR/Majority/Minority gates. GEMNIF and GEMFIC of the NOT gate is a constant (i.e. 1). The GEMFIC equations for AND, NAND, OR, and NOR gates share a commonality in that they have a singleton ON-set or OFF-set. Similarly, the GEMFIC equations for XOR, XNOR, Majority, and Minority gates are the same since they have equal ON-set and OFF-set cardinalities [47], [48].

$$\text{GEMFIC}_{\text{NOT}} = \frac{2}{2} = 1 \quad (4)$$

$$\text{GEMFIC}_{\text{AND/NAND/OR/NOR}} = \frac{2(2^n - 1)}{2^n(2^n - 1)} = 2^{1-n} \quad (5)$$

$$\text{GEMFIC}_{\text{XOR/XNOR/Majority/Minority}} = \frac{2^{2n-1}}{2^n(2^n - 1)} = \frac{2^{n-1}}{2^n - 1} \quad (6)$$

III. GEMNIF AND GEMFIC ESTIMATES FOR VARIOUS LOGIC GATE TYPES

Figures 2 and 3 show a plot of GEMNIF and GEMFIC estimates corresponding to various logic gate types assuming inputs ranging from 1 to 7. Modern digital standard cell libraries such as [46] do not feature AND, NAND, OR, and NOR gates with fan-in greater than 4, and XOR and XNOR gates with fan-in greater than 3. Also, only a 3-input majority/minority gate is normally included in most standard cell libraries, and higher fan-in majority/minority gates may be found in some digital cell libraries or may have to be custom-designed depending on the requirement. In Figures 2 and 3 however, we consider a wide range of inputs for the different gate types mainly to observe the impact that an increase in fan-in has on the GEMNIF and GEMFIC values. In this context, the lower the values of GEMNIF and GEMFIC the better is the fault tolerance, and the maximum values of GEMNIF and GEMFIC are integer 1.

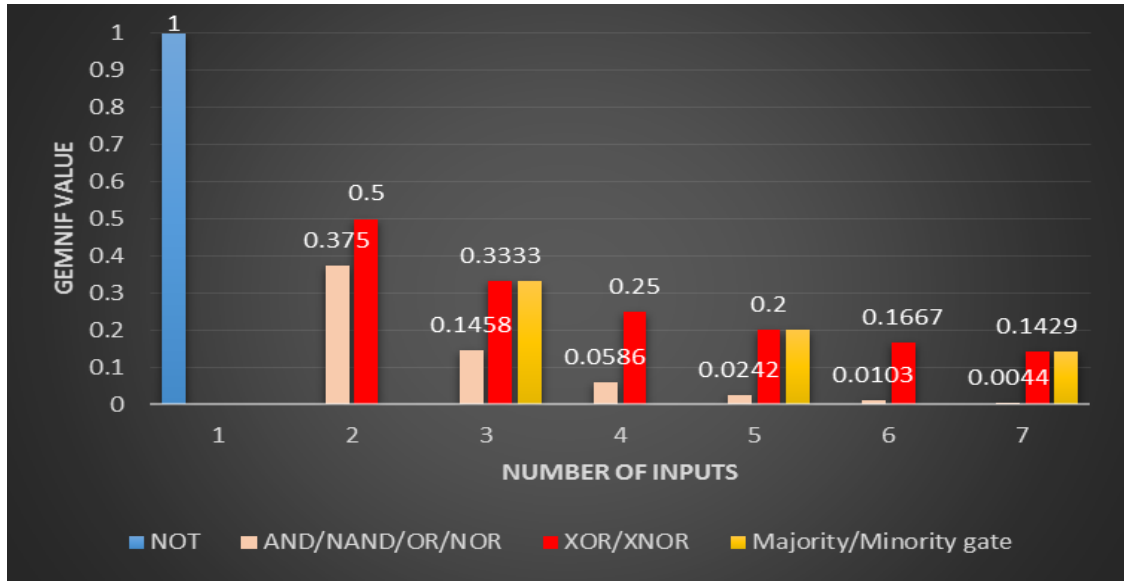

Figure 2. GEMNIF plots corresponding to various gate types. Although the GEMNIF values of XOR/XNOR and Majority/Minority gates are the same, the respective plots are explicitly shown above since the majority/minority gate has only an odd number of inputs. Hence, GEMNIF plots for the majority/minority gates will be applicable for those scenarios when the primary inputs are odd (i.e. 3, 5, 7 etc.). The NOT gate has a single input and a single output, and it has constant GEMNIF and GEMFIC values of 1. The values of $n = 1, 2, 3, \ldots$ are substituted into (3) and into the GEMNIF (i.e. GEM) equations given in [33] to obtain the above plots.

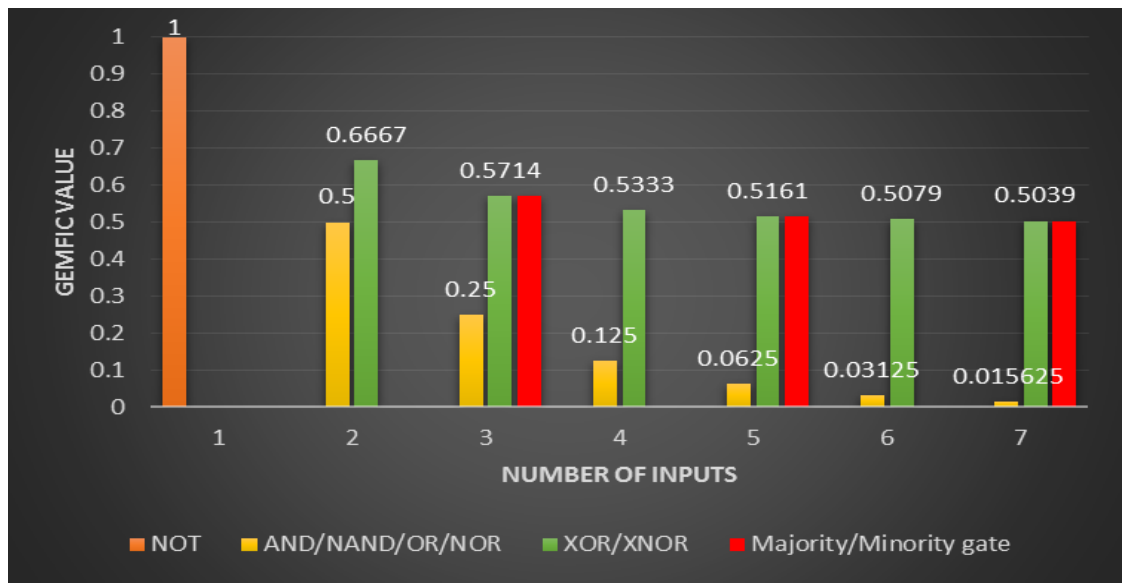

Figure 3. Figure 3. GEMFIC plots corresponding to various gate types. The values of $n = 1, 2, 3, \ldots$ are substituted into (4), (5) and (6) to obtain the above plots.

Regardless of the metric adopted i.e. whether GEMNIF or GEMFIC, there are some common trends noticed in Figures 2 and 3. With an increase of fan-in i.e. the number of inputs, both the singleton (AND/NAND/OR/NOR) and the equivalent (XOR/XNOR/Majority/Minority) logic gate types exhibit decreases in the GEMNIF and GEMFIC values. For example, the 2-input AND gate has a GEMNIF value of 0.375, and the 3-input AND gate has 61.1% reduced GEMNIF value. The GEMFIC value of the 2-input AND gate is 0.5, and the GEMFIC value of the 3-input AND gate is just 0.25, which implies a 50% reduction for the latter. Now considering the majority function, the 3-input majority gate has a GEMNIF value of 0.3333 and a GEMFIC value of 0.5714. In relative comparison, the 5-input majority gate has a 40% reduced GEMNIF value which is 0.25, and a 10% reduced GEMFIC value which is 0.5161. Thus, an obvious way to potentially improve the fault tolerance of a logic gate, when it may be subject to fault(s) on its primary inputs due to transient or permanent effects is to increase its fan-in.

GEMNIF evaluates the output errors based on the actual number of input fault occurrences, whereas GEMFIC evaluates the output errors based on just the faulty input patterns. A faulty input pattern may consist of one or more input fault(s) and so a mere count of the faulty input patterns may be an underestimation of a gate's fault tolerance. This is possibly the reason why the GEMFIC values for the various logic gate types shown in Figure 3 are higher than those of the corresponding GEMNIF values. Hence, it may be advisable to use the GEMFIC as a preliminary parameter for a quick fault tolerance assessment as the computational complexity associated with the GEMFIC estimation may be less than that of GEMNIF though the former is likely to be computationally less accurate than the latter.

When we consider an *n*-input logic gate or function, there will be $2^n$ unique input patterns and for each input pattern there can be a maximum of $(2^n - 1)$ distinct faulty input patterns. Since, here, GEMFIC considers only the actual input patterns and the associated faulty input patterns, its computational complexity may be directly characterized as $O[2^n(2^n - 1)]$. On the other hand, since GEMNIF considers the actual number of input fault occurrences, the actual number of input faults in relation to the actual number of faulty input patterns should be calculated and based on a weighted-average, the correct GEMNIF estimate is to be obtained, which is quite cumbersome though.

To simplify this calculation however, if on average, each faulty input pattern encompasses approximately $\frac{n+1}{2}$ input faults, the computational complexity of GEMNIF may be quickly characterized as $O\left[\{2^n(2^n - 1)\} \times \frac{n+1}{2}\right]$ which is a likely approximation and hence may not be reliable. Hence, for a reliably accurate fault tolerance estimation, GEMNIF is preferable compared to GEMFIC although GEMFIC can be used at an early stage.

IV. CONCLUSION

This paper has presented a mathematical estimation and an analysis of the fault tolerance of various logic gate types when subject to potential fault(s) on their inputs through two generalized error metrics viz. GEMNIF and GEMFIC. As mentioned in [29] that not all gates created are equal, it is found that gate types with a singleton ON-set or OFF-set (AND, NAND, OR, NOR) exhibit greater fault tolerance i.e. have less GEMNIF and GEMFIC values when compared to gate types (XOR, XNOR, Majority, Minority) which feature identical ON-set and OFF-set cardinalities. Building upon a previous work [33], the mathematical expressions pertaining to GEMNIF and GEMFIC for the majority/minority gates were deduced, and the mathematical expressions for GEMFIC corresponding to NOT, AND, NAND, OR, and NOR gates were also deduced. The important findings are: i) both GEMNIF and GEMFIC portray similar fault tolerance properties with respect to the various logic gate types although GEMNIF is computationally more expensive and more accurate than GEMFIC, and that GEMFIC may serve as an early indicator, ii) GEMNIF and GEMFIC equations of majority, minority, XOR and XNOR gates are all similar, and iii) with an increase of fan-in, the fault tolerance of all the gate types tend to improve resulting in less output errors. As a future work, it is to be investigated how these findings can be effectively incorporated into the fault-tolerant design of digital electronic circuits and systems.